\documentclass[12pt]{article}
\usepackage{amsmath,amsthm,amsbsy,amsfonts}
\usepackage{bm,graphicx,psfrag,epsf}
\usepackage{enumerate}
\usepackage[round, authoryear]{natbib}
\usepackage{url} 

\usepackage{ifpdf}
\usepackage{bbm}
\usepackage[doublespacing]{setspace}
\usepackage[usenames]{color}
\RequirePackage[OT1]{fontenc}
\usepackage{algorithm}
\usepackage{algpseudocode}

\usepackage{paralist}

\usepackage{amsbsy}

\usepackage[mathscr]{eucal}

\usepackage{bm}

\usepackage{xspace}

\usepackage{subfig}

\mathchardef\mhyphen="2D

\ifpdf
\usepackage{xr-hyper}
\usepackage[colorlinks,urlcolor=blue,citecolor=blue,linkcolor=blue]{hyperref}
\else
\newcommand{\href}[2]{{#2}}
\usepackage{url}
\fi

\ifpdf
\newcommand{\Sec}[1]{\hyperref[sec:#1]{Section~\ref*{sec:#1}}} 
\newcommand{\App}[1]{\hyperref[sec:#1]{Appendix~\ref*{sec:#1}}} 
\newcommand{\Supp}[1]{\hyperref[sec:#1]{Supplement~\ref*{sec:#1}}} 
\newcommand{\Eqn}[1]{\hyperref[eq:#1]{{\rm (\ref*{eq:#1})}}} 
\newcommand{\Part}[1]{\hyperref[part:#1]{(\ref*{part:#1})}} 
\newcommand{\Fig}[1]{\hyperref[fig:#1]{Figure~\ref*{fig:#1}}} 
\newcommand{\Tab}[1]{\hyperref[tab:#1]{Table~\ref*{tab:#1}}} 
\newcommand{\Thm}[1]{\hyperref[thm:#1]{Theorem~\ref*{thm:#1}}} 
\newcommand{\Lem}[1]{\hyperref[lem:#1]{Lemma~\ref*{lem:#1}}} 
\newcommand{\Prop}[1]{\hyperref[prop:#1]{Proposition~\ref*{prop:#1}}} 
\newcommand{\Cor}[1]{\hyperref[cor:#1]{Corollary~\ref*{cor:#1}}} 
\newcommand{\Def}[1]{\hyperref[def:#1]{Definition~\ref*{def:#1}}} 
\newcommand{\Alg}[1]{\hyperref[alg:#1]{Algorithm~\ref*{alg:#1}}} 
\newcommand{\Ex}[1]{\hyperref[ex:#1]{Example~\ref*{ex:#1}}} 
\newcommand{\As}[1]{\hyperref[as:#1]{Assumption~{\rm\ref*{as:#1}}}} 
\newcommand{\Reg}[1]{\hyperref[as:#1]{Condition~\ref*{reg:#1}}} 
\newcommand{\AlgLine}[2]{\hyperref[alg:#1]{line~\ref*{line:#2} of Algorithm~\ref*{alg:#1}}}
\newcommand{\AlgLines}[3]{\hyperref[alg:#1]{lines~\ref*{line:#2}--\ref*{line:#3} of Algorithm~\ref*{alg:#1}}}
\else
\newcommand{\Sec}[1]{{Section~\ref{sec:#1}}} 
\newcommand{\App}[1]{{Appendix~\ref{sec:#1}}} 
\newcommand{\Supp}[1]{{Supplement~\ref{sec:#1}}} 
\newcommand{\Eqn}[1]{{(\ref{eq:#1})}} 
\newcommand{\Part}[1]{{(\ref{part:#1})}} 
\newcommand{\Fig}[1]{{Figure~\ref{fig:#1}}} 
\newcommand{\Tab}[1]{{Table~\ref{tab:#1}}} 
\newcommand{\Thm}[1]{{Theorem~\ref{thm:#1}}} 
\newcommand{\Lem}[1]{{Lemma~\ref{lem:#1}}} 
\newcommand{\Prop}[1]{{Proposition~\ref{prop:#1}}} 
\newcommand{\Cor}[1]{{Corollary~\ref{cor:#1}}} 
\newcommand{\Def}[1]{{Definition~\ref{def:#1}}} 
\newcommand{\Alg}[1]{{Algorithm~\ref{alg:#1}}} 
\newcommand{\Ex}[1]{{Example~\ref{ex:#1}}} 
\newcommand{\Reg}[1]{{R~\ref*{reg:#1}}} 
\fi

\newcommand{\argmin}{\operatorname*{argmin}}
\newcommand{\Real}{\mathbb{R}}

\newcommand{\T}{^{\sf T}} 

\newcommand{\V}[1]{{\bm{\mathbf{\MakeLowercase{#1}}}}} 

\newcommand{\M}[1]{{\bm{\mathbf{\MakeUppercase{#1}}}}} 

\newcommand{\amp}{\mathop{\:\:\,}\nolimits}

\newtheorem{proposition}{Proposition}[section]
\newtheorem{theorem}{Theorem}[section]

\usepackage[in]{fullpage}

\usepackage{amsfonts}
 \usepackage{cases}
\usepackage{multirow}
\usepackage{textcmds}

\begin{document}

\def\spacingset#1{\renewcommand{\baselinestretch}%
{#1}\small\normalsize} \spacingset{1}


  \title{\bf A Convex-Nonconvex Strategy for Grouped Variable Selection}
  \author{Xiaoqian Liu
     \\
    Department of Statistics, North Carolina State University, Raleigh, NC\\
     Aaron J. Molstad \\
    Department of Statistics, University of Florida, Gainesville, FL\\
    and \\
   Eric C. Chi \\
    Department of Statistics, Rice University, Houston, TX}
    
    \date{}
  \maketitle


\begin{abstract}
This paper deals with the grouped variable selection problem. A widely used strategy is to augment the negative log-likelihood function with a sparsity-promoting penalty.
 Existing methods include the group Lasso, group SCAD, and group MCP. The group Lasso solves a convex optimization problem but is plagued by underestimation bias. The group SCAD and group MCP avoid this estimation bias but require solving a nonconvex optimization problem that may be plagued by suboptimal local optima. In this work, we propose an alternative method based on the generalized minimax concave (GMC) penalty, which is a folded concave penalty that maintains the convexity of the objective function. We develop a new method for grouped variable selection in linear regression, the group GMC, that generalizes the strategy of the original GMC estimator. We present an efficient algorithm  for computing the group GMC estimator and also prove properties of the solution path to guide its numerical computation and tuning parameter selection in practice. We establish error bounds for both the group GMC and original GMC estimators. A rich set of simulation studies and a real data application indicate that the proposed group GMC approach outperforms  existing  methods  in several different aspects under a wide array of scenarios. 
\end{abstract}

\noindent%
\vfill

\newpage
\spacingset{1.45} 


\section{Introduction}


Consider the classical linear regression setting where the data  have been generated according to the following model:
\begin{eqnarray}
\label{linear}
\V y &  = & \M X \V \beta^\star+\V \epsilon,
\end{eqnarray}
where $\V y\in \Real^n$ is a response vector, $\M X \in \Real^{n\times p}$ is a fixed design matrix whose columns are $p$
covariate variables, and $\V \epsilon$ is a vector of independent noise variables with mean zero and variance $\sigma^2$. 
 In modern statistical applications, we often have $p \gg n$ where the ordinary least squares estimator is not well defined. 
 
 A natural strategy to address this issue is to assume that $\V \beta^\star$ is sparse so as to improve both the prediction accuracy and the interpretation of the model. To that end, \cite{Lasso} developed the least absolute shrinkage and selection operator (Lasso) which performs both coefficient estimation and variable selection. The Lasso estimator is a solution to 
\begin{eqnarray*}
\label{Lasso}
   \operatorname*{minimize}_{{\V \beta} \in \Real^p} \frac{1}{2n}\lVert \V y-\M X \V \beta\rVert_2^2+\lambda \lVert \V \beta \rVert_1,
\end{eqnarray*}
where the objective function is a sum of the squared error loss, which represents the lack-of-fit,  and the $l_1$-norm penalty, which encourages sparsity in the estimated model. The $l_1$-norm is able to perform variable selection because of its singularity at the origin. The nonnegative tuning parameter $\lambda$ balances the trade-off between the goodness-of-fit and the model complexity.  

The Lasso  is one of the most popular penalized regression formulations for selecting individual variables but cannot immediately deal with certain types of structured sparsity. For example, in many statistical applications, variables may have a natural group structure. A classic example in regression is the encoding of a single categorical variable using a group of dummy variables. In such case, what we need is a method for selecting, or not selecting, the entire set of dummy variables. The most prominent work in grouped variable selection is the group Lasso \citep{grouplasso}, which is a natural extension of Lasso  and solves the following penalized least squares problem:
\begin{align}
\label{gLasso}
   \operatorname*{minimize}_{{\V \beta} \in \Real^p} \frac{1}{2n}\lVert  \V y-\sum_{j=1}^J \M  X_{\cdot,j}  \V \beta_j\lVert_2^2+\lambda \sum_{j=1}^J K_j \lVert \V \beta_j\lVert_2,
\end{align}
where the $p$ covariates are divided into $J$ groups, $ \V \beta=(\V \beta_1\T,...,  \V \beta_J\T )\T \in \Real^p$ with $ \V \beta_j \in \Real^{p_j}$ and $\sum_{j=1}^J p_j=p$. The matrix $\M X_{\cdot, j}$ is the  submatrix of $\M X$ whose columns correspond to the variables in the $j$-th group. The $K_j$'s are nonnegative weights used to adjust for the group sizes. A typical choice of $K_j$ is  $\sqrt{p_j}$. The group Lasso employs the $l_2$-norm of the group coefficients as a component of the penalty function. One can also view the penalty as applying the $l_1$-norm to the vector of $l_2$-norms of the groups, which enforces  sparsity at the group level while encouraging ridge regression-like shrinkage within a group.

Since the introduction of the group Lasso, many variants and generalizations have been proposed and investigated. 
 \cite{Kim2006} designed a blockwise sparse regression method to extend the idea of the group Lasso to general loss functions but used the same  penalty as the group Lasso.
 \cite{meier2008} derived the group Lasso for logistic regression and presented an efficient algorithm for fitting related generalized linear models.
  \cite{CAP} developed a family of composite absolute penalties  for grouped and hierarchical variable selection, which includes the group Lasso as a special case. 
 \cite{adgroup} generalized the adaptive Lasso  to an  adaptive group Lasso method to improve the variable selection performance.
 \cite{sparsegroup} introduced
 the sparse-group Lasso method which carries out both  individual and grouped variable  selection, also known as  bi-level variable selection, by incorporating a combination of the $l_1$-norm and the $l_2$-norm into the penalty. The grouping information included in the models discussed above has led to  improvements in both estimation accuracy and model interpretability. Many applications can be found in the corresponding references.

Nonetheless, despite its many desirable characteristics, the group Lasso and its variants suffer from the same drawback as the Lasso, namely, they tend to underestimate large magnitude coefficients due to applying the same amount of shrinkage on all coefficients. Nonconvex penalties, such as the smoothly clipped absolute deviation (SCAD) \citep{SCAD}  and the minimax concave
penalty (MCP) \citep{MCP}, have been developed as alternatives to the Lasso that can diminish the estimation bias in the Lasso estimator.
By applying such nonconvex penalties to the $l_2$-norms of the group coefficients, it is natural to obtain grouped variable selection via nonconvex penalization, such as the group  SCAD and group MCP \citep{groupSCAD1, groupSCAD2, Huang2012}. On the other hand, applying the nonconvex penalties to a vector of the $l_1$-norm of the groups can achieve bi-level variable selection and  better estimation of coefficients. Available methods  include the group bridge approach \citep{Huang2009}, the composite penalization methods  \citep{Breheny2009, Huang2012},  and the group exponential Lasso  \citep{Breheny2015b}.

Grouped variable selection models with nonconvex penalties are not without their disadvantages, however. The nonconvex penalty, though beneficial for the estimation of coefficients, leads to a nonconvex optimization problem. Objective functions in nonconvex programming typically possess multiple local optima which are not global optima, and algorithms for solving nonconvex optimization problems such as gradient or coordinate descent may be  trapped in local optima.
Early work on statistical theory for SCAD or MCP penalized least squares estimators focused on either error bounds for global optima \citep{zhang2012general} or local optima obtained through specific initialization schemes and algorithms \citep{fan2014strong}. 
More recently, \citet{loh2015regularized} established statistical properties which apply to all stationary points of SCAD or MCP penalized least squares objective functions (though their results do not apply directly to group SCAD or MCP penalized estimators). However, empirical results from \citet{fan2014strong} and \citet{loh2015regularized}, among others, suggest that in practice some stationary points perform much better than others, especially when the overall objective function is highly nonconvex, e.g., see the Remark on $(\alpha_1, \mu)$ and Figure 4 of \citet{loh2015regularized}. 

To overcome the drawbacks of 
nonconvex optimization, one line  of research, commonly referred to as the convex-nonconvex strategy, has been studied in the field of signal processing. 
This strategy adopts the so-called convexity-preserving nonconvex penalization, namely, the penalties are nonconvex but capable of maintaining the convexity of  the whole objective function. The idea of convexity-preserving nonconvex penalties was introduced by \cite{Blake1987}, \cite{Nikolova1998}, and \cite{Nikolova2010}, and then further investigated in \cite{Bayram2015}, 
\citet{Selesnick2017}, and \cite{Zou2018}. In particular, \cite{GMC} proposed a novel  nonconvex penalty function for the regularized least squares problem, which they call  the generalized minimax concave (GMC) penalty. The GMC penalty is given by 
\begin{equation}
\label{GMC}
    \psi_{\M  B}( \V \beta) \amp = \amp\| \V \beta \|_{1}-\min _{ \V v \in \Real^{p}}\left\{\| \V v\|_{1}+\frac{1}{2}\| \M B( \V \beta- \V v)\|_{2}^{2}\right\},
\end{equation}
which can guarantee the convexity of the optimization problem under a suitable condition on the matrix parameter $\M B$. \cite{Lanza2019} introduced a more general parametric nonconvex nonseparable regularizer  for a convex nonconvex variational model.
\cite{LIGMC} extended the idea of the GMC to a linearly involved
GMC penalty, which is applicable to more general situations, especially for recovering piecewise constant signals. 
 \cite{Liu2020} further investigated the linearly
involved GMC penalty, proposing a new method for choosing the matrix parameter $\M B$ in the penalty and
providing an additional algorithm to compute the solution path of the corresponding penalized least squares problem. The convex-nonconvex strategy bridges the gap between convex and
nonconvex approaches in prior work. The nonconvex penalty can greatly mitigate
the estimation bias for large magnitude coefficients, while the convexity of the optimization problem
guarantees that all local minima are global minima and opens the door to using  many efficient algorithms that are available for convex optimization problems. 

In this paper, we focus on  grouped  variable selection in linear regression and propose a new generalization of the GMC pen{}alty, called the group GMC penalty, which is also a convexity-preserving nonconvex penalty. We present the convexity-preserving condition for the group GMC model and some properties of its solution path. 
To solve the proposed optimization
problem, we cast it as a saddle-point problem and provide a primal-dual algorithm for iteratively computing its saddle point. Theoretically, we establish  error bounds for both the group GMC penalized least squares estimator and, as a special case, the GMC  estimator of  \cite{GMC}, which to the best of our knowledge have not been established yet.  In contrast to the theory for SCAD or MCP penalized least squares estimators, our theory applies only to global minimizers, which our algorithm is guaranteed to obtain.
We evaluate the
effectiveness of the proposed approach by comparing it with existing  grouped variable selection methods in several different simulation experiments and a real data  application.

The rest of this paper is organized as follows.  In Section \ref{sec2}, we first review the GMC penalty and its  relation to existing folded concave penalties. Then we formulate the group GMC  penalty and the corresponding optimization problem for grouped variable selection in linear regression.  We also include the convexity-preserving condition and some theoretical properties of the solution path  in this section.   In Section \ref{sec3}, we present in detail how to solve the proposed optimization problem with a first-order primal-dual algorithm.
In Section \ref{sec4}, we study the statistical properties of both the group GMC estimator and the original GMC estimator by establishing  $l_2$-norm error bounds.
In Sections  \ref{sec5} and \ref{sec6},
we  report on  numerical experiments and a real data application.  We close  with a discussion in Section \ref{sec7}. All proofs in this paper are included in the supplement.

\section{Group GMC}
\label{sec2}

\subsection{GMC and MCP}

 We  first review  the GMC penalty to help readers understand the relationship between the GMC and the MCP in \cite{MCP}. For that purpose, we have to recall the definition of the infimal convolution and the Huber function. 
 
 The infimal convolution of two functions $f$ and $g$  is 
   \begin{eqnarray*}
   \label{infimal}
     (f \Box g)( \V \beta) & = &
     \inf_{\V v \in \Real^p} \left\{ f(\V v) + g( \V\beta - \V v)\right\}.
  \end{eqnarray*}
 The Huber function  \citep{Huber} is defined as
  \begin{eqnarray*}
     \label{Huber}
     h(\beta) & = & \begin{cases}
     \frac{1}{2}\beta^2 & \text{if $\lvert \beta \rvert \leq 1$} \\
     \lvert \beta \rvert - \frac{1}{2} & \text{if $\lvert \beta \rvert > 1$}
     \end{cases},
 \end{eqnarray*}
which can be equivalently expressed as the infimal convolution
   \begin{eqnarray*}
     \label{Huber2}
     h(\beta) & = & \min_{v \in \Real} \left\{ |v|+\frac{1}{2}(\beta-v)^2 \right\}, 
 \end{eqnarray*}
 where the infimum is replaced by a minimum since the infimum in the definition is attained.
 
\cite{GMC} defined the scaled version of the Huber function as
   \begin{eqnarray*}
     h_b(\beta) & = & h(b^2\beta)/b^2 \amp = \amp \min_{v \in \Real} \left\{ |v|+\frac{1}{2}b^2(\beta-v)^2 \right\},
     \label{sHuber}
 \end{eqnarray*}
 where $b \neq 0$ is a scalar parameter. In the special case where $b=0$,  $h_b(\beta) = 0$. 
Based on the scaled Huber function,  the scaled minimax concave (MC) penalty is given by
   \begin{eqnarray}
     \label{smcp}
        \phi_b(\beta) & = & |\beta|-h_b(\beta).
 \end{eqnarray}

 After defining the scaled Huber function in the univariate case, \cite{GMC} proposed a natural multivariate generalization.
 Given a matrix parameter $\M B$, the generalized Huber function $H_{ \M B}: \Real^p \to \Real$  is written as
    \begin{eqnarray*}
     H_{ \M B}(\V \beta) & = & \inf_{\V v \in \Real^p} \left\{ \lVert \V v\lVert_1 + \frac{1}{2}\lVert \M B ( \V\beta-  \V v)\lVert_2^2\right\},
 \end{eqnarray*}
which is a convex function; the infimum is attained since the $l_1$-norm is coercive.  
 Mimimicking the univariate scaled minimax MC penalty, the generalized MC (GMC) penalty is defined as the difference of the $l_1$-norm and the generalized Huber function.
    \begin{eqnarray}
     \label{gmc}
     \psi_{ \M B}( \V \beta) & = & \lVert  \V \beta \lVert_1 - H_{ \M B}(\V \beta),
 \end{eqnarray}
 which coincides with \eqref{GMC}. Since the difference of two convex functions is not necessarily convex, the GMC penalty function is in general nonconvex. But  as mentioned in the introduction, the GMC penalty can maintain the convexity of the penalized least squares problem by suitably choosing $\M B$. Details can be found in \cite{GMC}.
 
 Recall that the MCP defined in \cite{MCP} is expressed as 
 \begin{eqnarray}
 \label{MCP}
 P_{\lambda, \gamma}(\V \beta) & = & \sum_{j=1}^p \rho_{\lambda, \gamma}(|\beta_j|),
 \end{eqnarray}
 where the univariate MCP function defined on $[0, \infty)$ is
 \begin{eqnarray}
      \label{mcp}
     \rho_{\lambda, \gamma}(|\beta_j|) & = & \left\{\begin{array}{ll}\lambda |\beta_j|-\frac{\beta_j^{2}}{2 \gamma}, & \text { if } |\beta_j| \leq \gamma \lambda \\ \frac{1}{2} \gamma \lambda^{2}, & \text { if } |\beta_j|>\gamma \lambda\end{array}\right.
 \end{eqnarray}
 where $\lambda \geq 0$ is the tuning parameter  controlling the degree of penalization, and $\gamma>1$ is a  hyper-parameter that determines the degree of concavity of the MCP.  The MCP function converges pointwise to the $l_1$-norm as $\gamma \to \infty$  and to the $l_0$-norm as $\gamma \to 1$, therefore the MCP provides a continuum of penalties by varying the value of $\gamma$.

Now let  us have a  closer look at the similarities and differences between the GMC penalty \eqref{gmc} and the MCP \eqref{MCP}. In the univariate case, the GMC penalty coincides with the scaled MC penalty \eqref{smcp}
and the MCP \eqref{MCP} reduces to $P_{\lambda, \gamma}(\beta)= \rho_{\lambda, \gamma}(|\beta|)$. If we set $b^2 = 1/\gamma\lambda$, then $\rho_{\lambda, \gamma}(|\beta|) = \lambda \phi_b(\beta)$.
 In other words,  \eqref{gmc} is equivalent to \eqref{MCP} up to a factor of $\lambda$. The difference between the  GMC penalty \eqref{gmc} and the MCP \eqref{MCP} lies  in how they are generalized from the univariate case to the multivariate one. The MCP \eqref{MCP}  takes an additive form from the univariate MCP function \eqref{mcp}, while the GMC penalty \eqref{gmc} is derived from the scaled MC penalty \eqref{smcp} via an infimal convolution, thus leading to a non-separable penalty function whenever $\M B\T \M B$ is non-diagonal.

The implications of expressing the MC penalty as an infimal convolution are non-trivial and lead to intrinsic differences with the standard MCP. It is well known that in the classic low-dimensional case where $n > p$, there exists a suitable hyper-parameter $\gamma$ choice for MCP that leads to a convex objective function but that no such $\gamma$ exists when $n < p$. In contrast, we will see that it is always possible to find a matrix $\M{B}$ that leads to a convex objective function for any $n$ and $p$. Thus, the GMC function enables the application of folded concave penalties in the high-dimensional case where $n \ll p$ without sacrificing convexity, opening the door to methods that can enjoy the best of both convex and nonconvex worlds.


\subsection{The group GMC model}

Based on the form of the GMC penalty \eqref{GMC} and mimicking the generalization from the Lasso to the group  Lasso, we define the group  GMC penalty as
\begin{align}
\label{grGMC}
    \Phi_{\M B}( \V \beta) \amp = \amp \sum_{j=1}^J K_j\| \V \beta_j\|_2-\min _{ \V v \in \Real^p}\left\{\sum_{j=1}^J K_j \| \V v_j\|_2+\frac{1}{2n}\| \M B( \V \beta- \V v)\|_{2}^{2}\right\},
\end{align}
where $\V \beta=( \V \beta_1\T,...,  \V \beta_J\T )\T \in \Real^p$ and $ \V v=( \V v_1\T,...,  \V v_J\T )\T \in \Real^p$. For each $j \in [J]$ where $[J]=\{1, \cdots, J\}$ for a positive integer $J$,  $\V \beta_j, \V v_j \in \Real^{p_j}$ with $\sum_{j=1}^J p_j=p$, and $K_j$ is the same as that in the group Lasso model \eqref{gLasso}. Here we insert a multiplier $1/n$ in the squared term of the group GMC penalty to put  it on the same scale with the squared error loss term in \eqref{gLasso}.

Therefore, the group GMC model for grouped variable selection and coefficient estimation in linear regression is cast  as the following optimization problem:
\begin{align}\label{grGMCmodel}
 \operatorname*{minimize}_{\V \beta \in \Real^p}~ \frac{1}{2n}\lVert \V y- \M  X  \V \beta \lVert_2^2+\lambda\Phi_{ \M B}( \V \beta),
\end{align}
where $\lVert  \V y- \M X\V \beta \lVert_2^2=\lVert \V y-\sum_{j=1}^J \M X_{\cdot, j}  \V \beta_j\lVert_2^2$ as  in \eqref{gLasso}. Here $\lambda \geq 0$ is again a tuning parameter that  controls the degree of penalization, while $\M B$ is a matrix parameter that controls the concavity of the group GMC penalty. Note that in our paper, we refer to $\lambda$ as the tuning parameter of the group GMC and treat the matrix $\M B$ as a hyper-parameter.

Similar to the GMC approach, the basic idea of the group GMC method is to maintain the convexity of the optimization problem while using a nonconvex penalty, which can be realized with an appropriate choice of the matrix hyper-parameter $\M B$.  The next proposition specifies the condition that  $\M B$ has  to satisfy to guarantee  the convexity of problem \eqref{grGMCmodel}. Recall that for two matrices $\M A$ and $\M B$,  $\M A \succeq \M B$ means  $\M A- \M B$ is positive semi-definite; similarly, $\M A \succ  \M B$ means  $\M A- \M B$ is positive definite.

\begin{proposition}
\label{prop1}
Let $\V y \in \Real^n, \M X \in \Real^{n \times p}$, and $\lambda \geq 0$. Define $F: \Real^p \to \Real$ as in \eqref{grGMCmodel}
\begin{eqnarray}
\label{Fgmc}
   F(\V \beta) & = & \frac{1}{2n}\lVert \V y- \M X \V \beta\lVert_2^2+\lambda\Phi_{\M B}(\V \beta) 
\end{eqnarray}
where $\Phi_{\M B} : \Real^p \to \Real$ is the group GMC penalty \eqref{grGMC}. If 
\begin{align}
\label{cd}
\M X\T  \M X \amp \succeq \amp \lambda \M B \T \M B,
\end{align}
then $F$ is a convex function. We call \eqref{cd} the  convexity-preserving condition for the group GMC problem \eqref{grGMCmodel}.
\end{proposition}

Note that the convexity-preserving condition \eqref{cd} can hold without any restriction on the problem dimension $p$ and the sample size $n$, namely, it can hold for both the low-dimensional case ($n \geq p$) and the high-dimensional case ($n<p$). To satisfy the convexity-preserving condition \eqref{cd}, an intuitive and simple choice for $\M B$ is 
\begin{eqnarray}
\label{B-eqn}
    \M B & = & \sqrt{\alpha/\lambda}\M X, \text{~~} 0 \leq \alpha \leq  1.
\end{eqnarray}
We refer to $\alpha$ as the convexity-preserving parameter of the group GMC  model since $\alpha$  controls the nonconvexity of the group GMC penalty. Setting $\alpha=0$ reduces the group GMC penalty
to the group Lasso penalty. And setting $\alpha=1$ gives a maximally nonconvex penalty which can maintain the convexity of the optimization problem \eqref{grGMCmodel}.
The convexity-preserving parameter 
$\alpha$ is another hyper-parameter of the group GMC method and needs to be chosen by users. We recommend a range of $ 0.4 < \alpha <1$ based on our simulation studies in Section \ref{sec5}.

The following proposition establishes the relationship between the group GMC and group MCP. It also clarifies the relationship between the GMC and MCP as a by-product.

\begin{proposition}
\label{gmc-mcp}
The group GMC method is equivalent to the group MCP method when $\M B \T \M B$ is diagonal and the diagonal elements are suitably designed. This equivalence also holds for the GMC and MCP.
\end{proposition}

We write the group GMC estimator, namely a minimizer to problem \eqref{grGMCmodel}, as $\hat{\V \beta}(\lambda)$ which explicitly represents the dependency of the solution  to \eqref{grGMCmodel} on the tuning parameter $\lambda$. We next discuss two properties of the solution path $\hat{\V \beta}(\lambda)$, that expedite the  numerical computation in practice. 

\begin{theorem}
\label{Th1}
Suppose $\M X\T \M X  \succ \lambda \M B\T \M B$, then the solution path $\hat{\V \beta}(\lambda)$ to the group GMC problem \eqref{grGMCmodel}   exists, is unique, and is continuous in $\lambda$.
\end{theorem}

Theorem \ref{Th1} tells us that the optimization problem \eqref{grGMCmodel} is well-posed.  Moreover, continuity of $\hat{\V \beta}(\lambda)$ opens the door to a homotopy strategy to reduce computation time when solving a sequence of problems over a grid of $\lambda$ values. Namely, we use the solution to the problem at the previous value of $\lambda$ to initialize, or warm start, the  next iterate for computing the solution at the next $\lambda$ value.

Intuitively, we may expect that all groups are excluded from the model when the tuning parameter $\lambda$ is sufficiently large. The following theorem confirms our intuition.

\begin{theorem}
\label{Th2}
The group GMC problem \eqref{grGMCmodel} has a unique solution $\hat{\V \beta}(\lambda)=\V 0_p$ for all $\lambda$ greater than  $\lambda_0 = \max_j  \left\{\lVert (\M X_{\cdot, j})\T \V y\rVert_2/(n K_j)\right\}$, where $\M X_{\cdot, j}$ and $K_j$ are as defined in \eqref{gLasso} for $j \in [J]$.
\end{theorem}

This second property is practically useful since it gives a range of $\lambda$, $[0, \lambda_0]$, to sample the full dynamic range of group sparse models, and as an added benefit the computation of $\lambda_0$ is straightforward.

We close this section with a few remarks. First, the group GMC penalty \eqref{grGMC}  depends on $\M B\T \M B$, not  $\M B$ itself. Therefore, there is no need to express $\M B$ explicitly when computing the solution path $\hat{\V \beta}(\lambda)$.  Second,  the two properties of the solution path hold for any matrix $\M B$ satisfying the convexity-preserving condition and are independent of how  $\hat{\V \beta}(\lambda)$ is computed, as they 
are intrinsic to the group GMC problem. Finally, Theorem~\ref{Th1} applies only in the classic setting where $n > p$. This is a more stringent condition than what is required to ensure the uniqueness of the Lasso \citep{Tibshirani2013}. The proof of the uniqueness of the Lasso solution hinged on the Karush-Kuhn-Tucker (KKT) conditions of the Lasso optimization problem. The KKT conditions for the group GMC problem, however, are more complicated than the KKT conditions for the Lasso. Generalization of the proof used in the Lasso case to the group GMC is not straightforward due to the more complicated KKT conditions of the latter. Nonetheless, we conjecture that relaxed conditions similar to those that ensure the uniqueness of the Lasso solution can be established and leave establishing these conditions for future work.


\section{Algorithms}
\label{sec3}

\subsection{Algorithm for the group GMC model}
In this subsection, we focus on the computation of the solution path $\hat{\V \beta}(\lambda)$ to the group GMC model \eqref{grGMCmodel}.
We first present a first-order primal-dual method, called the Primal-Dual Hybrid Gradient (PDHG) algorithm \citep{PDHG-1, PDHG-2}, for computing the solution to  non-smooth saddle-point problems. Then we formulate problem \eqref{grGMCmodel} as such a saddle-point problem, thus solving it by the PDHG algorithm.
 
 The PDHG method, also known as the Chambolle-Pock method, is widely used to solve the following saddle-point problem:
\begin{align}\label{PDHG}
\min_{ \V x \in \mathcal{X}} \max_{\V y \in \mathcal{Y}} f(\V x) + \V y\T\M A \V x - g(\V y)
\end{align}
where $f$ and $g$ are convex functions, $\M A \in \Real^{M \times N}$ is a matrix, and $\mathcal{X} \subset \Real^N$ and $\mathcal{Y} \subset \Real^M$ are convex sets.
A wide range of problems in statistics and machine learning can be cast as a case of \eqref{PDHG}, such as the scaled Lasso and total variation denoising. 

Algorithm \ref{PDHG-Basic} summarizes the basic PDHG steps for problem \eqref{PDHG}, where $\sigma_k$ and $\tau_k$ are stepsize parameters for updating  $\V x$ and  $\V y$, respectively. 
One can choose constant stepsizes, $\tau_k = \tau$ and $\sigma_k = \sigma$ with $\tau \sigma < \lVert \M A\T \M A\lVert^{-1}$, to guarantee the convergence of the PDHG algorithm.
Note that we use $\|\M A\|$ to denote the spectral norm of a matrix $\M A$.

\begin{algorithm}
\caption{Basic PDHG steps for problem \eqref{PDHG}}
\label{PDHG-Basic}
 Set $\V x_0 \in \Real^N, \V y_0 \in \Real^M, \sigma_k>0, \tau_k>0$
 \begin{algorithmic}[1]
 \setcounter{ALG@line}{0}
	  \Repeat
	 \State $\hat{\V x}_{k+1}= \V x_k-\tau_k \M A\T \V y_k$
	 \State $\V x_{k+1}= \operatorname*{arg\,min}_{ \V x \in \Real^N} f( \V x)+\frac{1}{2\tau_k}\lVert \V x-\hat{ \V x}_{k+1}\rVert_2^2$
	 \State $\hat{\V y}_{k+1}=\V y_k+\sigma_k \M A (2 \V x_{k+1}- \V x_k)$
	 \State $\V y_{k+1}= \operatorname*{arg\,min}_{ \V y\in \Real^M} g( \V y)+\frac{1}{2\sigma_k}\lVert  \V y-\hat{\V y}_{k+1}\rVert_2^2$
	 \Until{convergence}
\end{algorithmic} 
\end{algorithm}

We now recast the optimization problem \eqref{grGMCmodel} as a saddle-point problem 
\begin{align}\label{saddle}
\min_{ \V \beta \in \Real^p} \max_{ \V v \in \Real^p} f(\V \beta) + \V v\T \M Z \V \beta - g(\V v)
\end{align}
where 
\begin{align}\label{f}
   f(\V \beta) =  \frac{1}{2n}\lVert \V y- \M X \V \beta\lVert_2^2 +\lambda \sum_{j=1}^J K_j\| \V \beta_j\|_2-\frac{\lambda}{2n} \| \M B  \V \beta\|_2^2 ,
\end{align}
and 
\begin{align}\label{g}
  g(\V v) =  \frac{\lambda}{2n}\| \M B  \V v\|_2^2+\lambda \sum_{j=1}^J K_j \| \V v_j\|_2,
\end{align}
 and $\M Z =  \frac{\lambda}{n} \M B \T \M B \in \Real^{p \times p}$ is a symmetric matrix. In addition, both  \eqref{f} and \eqref{g} are convex functions under the convexity-preserving condition \eqref{cd}. 
 It is straightforward to see that problem $\eqref{saddle}$ is under the framework of  \eqref{PDHG} and thereby can be solved by the PDHG algorithm.

The basic PDHG method can be slow to converge, however.  In this paper, we implement the accelerated version, named the adaptive PDHG algorithm \citep{PDHG2013, PDHG2015}, to solve the
group GMC problem. 
We provide details about the adaptive PDHG for problem \eqref{saddle} and its convergence guarantees in the supplement.

\subsection{Algorithm for the PDHG updates}

The PDHG algorithm for solving the group GMC problem requires solving two subordinate optimization problems for updating  $\V \beta_{k+1}$ and $\V v_{k+1}$ which we describe next.

We first introduce an efficient algorithm, Fast Adaptive Shrinkage/Thresholding Algorithm (FASTA) \citep{FASTA1, FASTA2}, for solving optimization problems of the form
\begin{align}
\label{FASTA-prob}
    \underset{\V x \in \Real^N}{\text{minimize}}~ m(\V x) + h(\V x)
\end{align}
where $m$ is convex and Lipschitz differentiable, $h$ is proper, lower semi-continuous and convex, and $m+h$ is coercive. 
FASTA provides a simple framework for implementing the forward-backward splitting (FBS) method, also known as the proximal gradient method \citep{FBS}, to efficiently compute the solution to problem \eqref{FASTA-prob}.
Problems under the framework of \eqref{FASTA-prob} include the Lasso, noisy matrix completion,  and many other  regularized regression problems. 

Algorithm \ref{FASTA-Basic}
shows pseudocode of the basic FBS steps in FASTA for solving \eqref{FASTA-prob},  where $t_k$ is a positive stepsize parameter  and plays an important role in the convergence rate of the algorithm. The proximal operator of $h$ is given by
\begin{align}
\label{prox-h}
\operatorname{prox}_{h}(\V u)=\operatorname*{argmin}_{\V x \in \Real^N}\left(h(\V x)+\frac{1}{2}\lVert \V x-\V u\lVert_{2}^{2}\right).
\end{align}
The proximal operator esixts and is unique if $h$ is convex and lower semi-continuous. The key computation in FBS is the proximal mapping, and many regularizers $h$ in sparse learning admit proximal operators which  either have an explicit formula or can be evaluated by an efficient algorithm. For instance, the proximal operator of the $l_2$-norm can be explicitly expressed as 
\begin{equation*}
    \operatorname{prox}_{\lambda \lVert \cdot \lVert_2}(\V u)= \left(1-\frac{\lambda}{\lVert \V u\lVert_2}\right)_+ \V u,
\end{equation*}
where $(\cdot )_+ = \max\{0, \cdot\}$. We will use this expression in our proofs. The efficiency of computing the proximal operators makes the FBS method popularly used in practice.

\begin{algorithm}
\caption{Basic FBS  steps in FASTA for problem \eqref{FASTA-prob}}
\label{FASTA-Basic}
Set $\V x_0 \in \Real^N, t_k>0$
\begin{algorithmic}[1]
 \setcounter{ALG@line}{0}
	  \Repeat
	 \State $\hat{\V x}_{k+1}= \V x_k-t_k \nabla m(\V x_k)$
	\State $\V x_{k+1}=\operatorname*{prox}_{t_k h} (\hat{\V x}_{k+1})= \operatorname*{argmin}_{\V x \in \Real^N} t_k h(\V x)+ \frac{1}{2}\lVert  \V x-\hat{ \V x}_{k+1}\rVert_2^2$
	 \Until{convergence}
\end{algorithmic}
\end{algorithm}

We now rewrite the two  optimization problems for updating  $\V \beta_{k+1}$  and $\V v_{k+1}$ in the PDHG updates in the form  of \eqref{FASTA-prob}. First, the optimization problem for updating $\V \beta_{k+1}$ can be written as 

    \begin{align}
    \V \beta_{k+1} &=\operatorname*{argmin}_{ \V \beta \in \Real^p} f( \V \beta)+\frac{1}{2\tau_k}\lVert  \V \beta-\hat{\V \beta}_{k+1}\rVert_2^2 \notag\\
    & =\operatorname*{argmin}_{\V\beta \in \Real^p} \Big\{\frac{1}{2n}\lVert \V y- \M X\V\beta\lVert_2^2 -\frac{\lambda}{2n} \| \M B \V\beta\|_2^2 + \frac{1}{2\tau_k}\lVert \V\beta-\hat{\V\beta}_{k+1}\rVert_2^2\Big\} +\lambda \sum_{j=1}^J K_j\| \V \beta_j\|_2\label{update-beta}
\end{align}
Similarly, we  write  the optimization problem for updating $\V v_{k+1}$ as
\begin{align}
     \V v_{k+1}& =\operatorname*{argmin}_{ \V v\in \Real^p} g( \V v)+\frac{1}{2\sigma_k}\lVert  \V v-\hat{\V v}_{k+1}\rVert_2^2 \notag\\
     & = \operatorname*{argmin}_{ \V v\in \Real^p} \Big\{\frac{\lambda}{2n}\| \M B  \V v\|_2^2 +\frac{1}{2\sigma_k}\lVert  \V v-\hat{\V v}_{k+1}\rVert_2^2  \Big\}+\lambda \sum_{j=1}^J K_j \| \V v_j\|_2\label{update-v}
\end{align}

Both  \eqref{update-beta} and \eqref{update-v}
satisfy  the conditions on $m$ and $h$ in \eqref{FASTA-prob}. Therefore, we can  compute $\V\beta_{k+1}$ and $\V v_{k+1}$ by using Algorithm \ref{FASTA-Basic}.

One of the primary difficulties with FBS is that users must carefully choose the stepsize. Fortunately, many variants of FBS are available in FASTA for adaptively choosing stepsize and accelerating convergence. In this paper, we use the strategies adopted in the R package \texttt{fasta} to implement FASTA to get the solutions to \eqref{update-beta} and \eqref{update-v}.



\section{Statistical properties}
\label{sec4}
\subsection{Main results}
In this section, we consider the statistical properties of the group GMC estimator obtained by solving \eqref{grGMCmodel}. First, we demonstrate that the group GMC estimator achieves an error bound of the same asymptotic order as existing estimators. Second, we discuss how the choice of $\M B$, or $\M B \T \M B$ to be exact, affects the error bound. We will also contrast our assumptions and error bounds with $\M B = \sqrt{\alpha/\lambda} \M X$ to those under the group Lasso penalization.

We now define a number of important quantities. First, define 
\begin{equation} 
\label{eq:v_star}
\V v^\star \amp = \amp \argmin_{\V v \in \mathbb{R}^p} \left\{ \sum_{j=1}^J K_j \|\V v_j\|_2 + \frac{1}{2n}\|\M B(\V \beta^\star - \V v)\|_2^2\right\},
\end{equation}
where $\V \beta^\star$ is the true vector of coefficients and $\V v_j$ has the same dimension as $\V \beta^\star_j$ for each $j \in [J]$. Implicitly, $\V v^\star$ is a function of $\M B, \V\beta^\star, n,$ and $K_j$: we avoid notation indicating this dependence for improved readability. We can then define the sets $\mathcal{S} = \{j: \|\V \beta_j^\star\|_2 \neq 0, j \in [J]\}$ and $\mathcal{S}^c = [J] \setminus \mathcal{S}$ and use $|\mathcal{S}|$ to denote the cardinality of $\mathcal{S}$. We also define
\begin{eqnarray*}
\nu_j & = & \left\{ \begin{array}{ll}
K_j + n^{-1} \|[\M B\T \M B]_{j,\cdot}(\V \beta^\star - \V v^\star)\|_2, &j \in \mathcal{S}\\
K_j - n^{-1} \|[\M B\T \M B]_{j,\cdot}(\V \beta^\star - \V v^\star)\|_2, &j \in \mathcal{S}^c\\
\end{array}\right.
\end{eqnarray*}
where $[\M B\T \M B]_{j,\cdot} \in \mathbb{R}^{p_j \times p}$ is the submatrix of $\M B\T \M B$ with rows corresponding to the indices of $\V \beta^\star$ defining the $j$-th group for each $j \in [J].$  Finally, let $\bar{\nu} = \max_{j \in \mathcal{S}} \nu_j$ and $\underline{\nu} = \min_{k \in \mathcal{S}^c} \nu_k$. Both $\bar{\nu}$ and $\underline{\nu}$ play important roles in our error bounds. In brief, our results indicate that a good choice of $\M B$ is one in which $\bar{\nu}$ is minimized and $\underline{\nu}$ is maximized, while also, the $\nu_j$ for $j \in \mathcal{S}$ are large and $\nu_k$ for $k \in \mathcal{S}^c$ are small. 
For the sake of illustration, we will derive closed form expressions for the $\nu_j$ under a particular choice of $\M B$ in the next subsection.

Our results require a number of conditions and assumptions. Our first condition is that the submatrices of $\M X$ satisfy a simple scaling condition \citep{Negahban2012unified}. Specifically, we assume that 
$ \|\M X_{\cdot, j}\| \leq \sqrt{n}$ for all $j \in [J]$ where $\|\cdot\|$ is the spectral norm. Such a condition was used in, for example, Corollary 4 of \citet{Negahban2012unified}. In the case that $p_j = 1$, this simplifies to the standard column-wise scaling condition that each column of $\M X$ has squared Euclidean norm no greater than $n$ (e.g., see Example 11.1 of \citet{hastie143statistical}).  We also assume the following:
\begin{itemize}
\item[] \textbf{A1.} (Subgaussian errors). The data are generated from \eqref{linear} where $\V\epsilon \in \mathbb{R}^n$ has independent entries which are each $\sigma$-subgaussian random variables for $0 < \sigma < \infty$. That is, $\mathbb{ E}(\epsilon_i)=0$ and for all $t \in \mathbb{R}$, $ \mathbb{E}\{\exp(t\epsilon_i)\} \leq  \exp(t^2 \sigma^2/2)$ for each $i \in [n].$

\item[] \textbf{A2.} (Convexity) The matrix $\M B$ is chosen so that $\M X\T  \M X \succeq \lambda \M B\T \M B$.

\item[] \textbf{A3.} (Sample size) The sample size $n$ is sufficiently large so that $\nu_k > 0$ for all $k \in \mathcal{S}^c.$
\end{itemize}
Finally, we require a version of the well-known restricted eigenvalue condition which depends on the design matrix $\M X$ and the matrix parameter $\M B$.
\begin{itemize}
\item[] \textbf{A4.} (Restricted eigenvalue condition) For a fixed $c > 1$, define 
\begin{eqnarray*}
  \mathbb{C}_n(\mathcal{S}, \nu, c) & = & \left\{\V \Delta \in \mathbb{R}^p: \V \Delta \neq \V 0,   \sum_{k \in \mathcal{S}^c} \left(\nu_k - \frac{\underline{\nu}}{c}\right)\|\V \Delta_{k}\|_2 \leq \sum_{j \in \mathcal{S}} \left(\nu_j + \frac{\underline{\nu}}{c}\right) \|\V \Delta_{j}\|_2 \right\},
\end{eqnarray*}
  where we use a single notation $\nu$ to indicate the dependency on $\bar\nu, \underline \nu$, and $\nu_j$ for $j \in [J]$.
 We assume there exists a constant $k > 0$ such that for all $n$ and $p$, 
 \begin{eqnarray*}
0 & < & k \amp \leq \amp \kappa_{\M B}(\mathcal{S}, c) \amp = \inf_{\V \Delta \in \mathbb{C}_n(\mathcal{S}, \nu, c)} \frac{\V \Delta\T(\M X\T \M X - \lambda \M B\T \M B)\V \Delta}{2n\|\V \Delta\|_2^2}.
\end{eqnarray*}
\end{itemize}
We will discuss the modified restricted eigenvalue condition after stating our main result, which we prove in the supplement.
\begin{theorem} \label{thm:groupGMC_bound}
(Error bound for group GMC)
Let $c > 1$ and $k_1> 0$ be fixed constants. If assumptions \textbf{A1}--\textbf{A4} hold and
\begin{eqnarray*}
\lambda & = & \frac{2c\sigma}{\underline\nu}\left(\max_{j \in [J]}\sqrt{\frac{p_j}{n}} + \sqrt{\frac{k_1 \log (J)}{n}}\right),
\end{eqnarray*}
then with probability at least $1 - 2 \exp(-2 k_1 \log(J)),$
\begin{eqnarray*}
\|\hat{\V\beta}(\lambda) - \V \beta^\star \|_2 & \leq & \frac{2 c \sigma}{\kappa_{\M B}(\mathcal{S},c)}\left(\frac{\bar\nu}{\underline\nu} + \frac{1}{c}\right)\left\{\left(\max_{j \in [J]}\sqrt{\frac{|\mathcal{S}|p_j}{n}}\right) + \sqrt{\frac{|\mathcal{S}|k_1 \log (J)}{n}}\right\},
\end{eqnarray*}
where $\hat{\V \beta}(\lambda)$ is the group GMC estimator obtained from \eqref{grGMCmodel}.
\end{theorem}
Perhaps unsurprisingly, the group GMC estimator achieves the same asymptotic error rate as the group Lasso penalized least squares estimator. Where an improvement over other convex estimators could be realized is through judicious choice of $\M B$ such that $\kappa_{\M B}(\mathcal{S},c)$ is large and $\bar\nu/\underline{\nu}$ is small. We discuss this further in the next subsection.

As a consequence of our proof technique, we also establish an error bound for the original GMC estimator in \cite{GMC}: this is a special case of the group GMC estimator with each group consisting of a single coefficient. This is the first error bound for the GMC estimator that we are aware of. 
\begin{theorem}\label{thm:GMC_bound} (Error bound for GMC)
Let $c > 1$ and $k_2 \in (0,1/2)$ be fixed constants. Let $p_j = 1$ for $j \in [p]$ so that $\mathcal{S} = \{j: \V \beta^\star_j \neq 0, j \in [p]\}$. If assumptions \textbf{A1}--\textbf{A4} hold and $\lambda = (c\sigma/\underline\nu)\sqrt{2 \log (p/k_2)/n},$ then with probability at least $1 -2 k_2$
\begin{eqnarray*}
\|\hat{\V \beta}(\lambda) - \V \beta^\star \|_2 & \leq & \frac{c \sigma}{\kappa_{\M B}(\mathcal{S},c)}\left(\frac{\bar\nu}{\underline\nu} + \frac{1}{c}\right)\sqrt{\frac{2|\mathcal{S}|\log (p/k_2)}{n}},
\end{eqnarray*}
where $\hat{\V \beta}(\lambda)$ is the corresponding GMC estimator.
\end{theorem}
Like the group-penalized version, the GMC penalized estimator achieves the same well-known asymptotic $\sqrt{|\mathcal{S}|\log p/n}$ rate as its $l_1$-norm penalized counterpart. However, like the group GMC penalized estimator, an improvement over Lasso in finite sample settings may be realized through an inflation of the restricted eigenvalue $\kappa_{\M B}(\mathcal{S},c)$, and through the role of the $\nu_j$'s.

\subsection{Additional insights}
The restricted eigenvalue $\kappa_{\M B}(\mathcal{S},c)$ in \textbf{A4} differs from the analogous condition under the $l_2$-norm group penalization, which may partly explain the difference in performance observed in Section \ref{sec5}. Specifically, to establish error bounds for the group Lasso analog of \eqref{grGMCmodel}, the corresponding restricted eigenvalue condition posits a lower bound on
$\inf_{\V \Delta \in \mathbb{D}_n(\mathcal{S}, c)} \frac{\V \Delta \T \M X\T \M X \V \Delta}{2n\|\V \Delta\|_2^2},$
where 
\begin{eqnarray*}
\mathbb{D}_n(\mathcal{S}, c) & = & \left\{\V \Delta \in \mathbb{R}^p: \V \Delta \neq \V 0,   \sum_{k \in \mathcal{S}^c} K_k\|\V \Delta_{k}\|_2 \leq \left(\frac{c + 1}{c-1}\right) \sum_{j\in \mathcal{S}} K_j \|\V \Delta_{j}\|_2 \right\},
\end{eqnarray*}
when the tuning parameter is chosen to be at least as large as $c \max_j \|\V \epsilon\T \M X_{\cdot,j}\|_2 /n$.
The difference between restricted eigenvalue conditions comes both in terms of the function the infimum is taken with respect to and in terms of the set over which the infimum is taken. For example, when we take $\M B = \sqrt{\alpha/\lambda} \M X$ for $\alpha \in (0,1)$, we see that 
\begin{eqnarray*}
\kappa_{\M B}(\mathcal{S}, c) & = &  (1 - \alpha) \left\{\inf_{\V \Delta \in \mathbb{C}_n(\mathcal{S}, \nu, c)}\frac{\V \Delta\T \M X\T\M X \V \Delta}{2n\|\V \Delta\|_2^2}\right\},
\end{eqnarray*}
 which would at first glance seem to imply a $(1-\alpha)$ factor decrease in the restricted eigenvalue relative to that for the group Lasso estimator. However, the benefit comes through the potential reduction in volume of the set $\mathbb{C}_n(\mathcal{S}, \nu, c)$ relative to $\mathbb{D}_n(\mathcal{S},c)$.  If, for example, each $\nu_j + \underline{\nu}/c \geq (c+1) K_j$ for $j \in \mathcal{S}$ and each $\nu_k - \underline{\nu}/c \leq (c-1) K_k$ for $k \in \mathcal{S}^c$, this would 
guarantee the set $\mathbb{C}_n(\mathcal{S}, \nu, c)
$ has volume no greater than $\mathbb{D}_n(\mathcal{S}, c)$. More generally,  if $\underline\nu$ and many $\nu_j$ for $j \in \mathcal{S}$ are large, one may expect the reduction in volume of $\mathbb{C}_n(\mathcal{S}, \nu, c)$ relative to $\mathbb{D}_n(\mathcal{S}, c)$ to lead to a larger restricted eigenvalue and in turn, an improved error bound. 
In addition, since the restricted eigenvalue condition \textbf{A4} depends on the user-specified matrix $\M B$, one may select $\M B$ such that this condition is more plausible than the analogous condition under the group Lasso penalization. The matrix $\M B$ also affects the error bound through the $\nu_j$, both through the modification of $\mathbb{C}_n$ and the ratio $\bar{\nu}/\underline{\nu}$. 

To get a sense of how the $\nu_j$'s depend on the choice of $\M B$, we focus on a special case. 
\begin{proposition}
\label{remark1}
Suppose $n > p$ and $\M X\T \M X \succ \M O_p$. Consider the choice of $\M B = \sqrt{\eta/\lambda} \M I_p$ where $\eta>0$ is the smallest eigenvalue of $\M X\T \M X$ so that \textbf{A2} holds. In this situation,
\begin{eqnarray*}
\V v^\star_j & = &  \left(1- \frac{n K_j \lambda}{\eta \|\V \beta^\star_j\|_2}\right)_+\V \beta^\star_j,~~~~ j \in \mathcal{S}
\end{eqnarray*}
and $\V v^\star_j = 0$ for $j \in \mathcal{S}^c.$
It thus follows that for $j \in [J]$
\begin{eqnarray*}
\nu_j & = & K_j + K_j \mathbbm{1}\{ \|\V \beta_j^\star\|_2 > n K_j \eta/\lambda\} + \frac{\eta}{n\lambda}\|\V\beta^\star_j\|_2 \mathbbm{1}\{ \|\V \beta_j^\star\|_2 \leq n K_j \eta/\lambda\},
\end{eqnarray*}
where $\mathbbm{1}(\cdot)$ is the indicator function. 
\end{proposition}

In addition to providing insight regarding the $\nu_j$, in light of Proposition \ref{gmc-mcp}, this result provides a new lens through which the group MCP can be viewed in the settings in which group MCP and group GMC are equivalent. Existing theory for group MCP is primarily concerned with the oracle property rather than, say,  error bounds. 

Crucially, Proposition \ref{remark1} implies $\nu_j = K_j$ for all $j \in \mathcal{S}^c$ in the considered scenario. This choice of $\M B$ is useful because $\nu_j = K_j$ for $j \in \mathcal{S}^c$ (whereas alternative choices of $\M B$ will yield $\nu_j < K_j$ for some $j \in \mathcal{S}^c$), and because it enables us to express $\V v^\star$ explicitly in terms of $\V \beta^\star.$ In this setting of Proposition \ref{remark1}, if each $K_j = 1$ and the $\|\V \beta_j^\star\|_2$ are sufficiently large, then  $(\nu_j + \underline\nu/c) =  (2 + 1/c)$ for $j \in \mathcal{S}$ and $(\nu_k - \underline{\nu}/c) = (c-1)/c$ for $k \in  \mathcal{S}^c$ so that $\mathbb{C}_n(\mathcal{S}, \nu, c)$ could be written in the same form as $\mathbb{D}_n(\mathcal{S},c)$ with $(c+1)/(c-1)$ replaced with $(2c + 1)/(c - 1)$ (i.e., $\mathbb{C}_n(\mathcal{S}, \nu, c)$ has less volume than $\mathbb{D}_n(\mathcal{S},c)$). 
 
In practice, of course, the $\nu_j$ cannot be computed since they depend on $\V \beta^\star$. Likewise, the $\M B$ which minimizes the bounds in Theorem \ref{thm:groupGMC_bound} also depends on $\V \beta^\star$ and the set $\mathcal{S}$, so this cannot be used in practice.

\section{Simulation studies}
\label{sec5}

We investigate the practical performance of the proposed group GMC method with experiments that build upon on the simulation scenarios in \cite{grouplasso}. We also compare the group GMC with  the group Lasso, group MCP, and group SCAD. The computation of the  three existing methods is done using the R package \texttt{grpreg} developed by \cite{Breheny2015}, and  the hyper-parameters in the group  MCP and group  SCAD are set as the default values given in the R package.

There are four linear regression models considered in the simulation study in \cite{grouplasso}. In this work, we consider the  two most complicated ones, an ANOVA model with all two-way interactions and an additive model with both categorical and continuous variables. More importantly,
we  study different cases for each model to explore the effects of interesting factors, including the signal-to-noise ratio (SNR), the correlation among groups, the problem dimension, and the convexity-preserving parameter (only for the group GMC). In each case, we run the experiment for $100$ replications and evaluate different methods with
respect to: (i) mean squared error (MSE) of the estimated coefficients; (ii) prediction error defined as $\lVert \M X\hat{\V \beta} -\M X\V\beta^\star \rVert_2^2/n$ where $n$ is the sample size, $\M X$ is the design matrix, and $\V \beta^\star$ and $\hat{\V \beta}$ are the vectors of true and estimated coefficients respectively; (iii)
support recovery with
respect to F1 score, number of true positives (TP), and number of false positives (FP). The F1 score is a metric of support recovery performance that accounts for both TP and FP; 
it is defined as $\text{F1} = 2\text{TP}/(2\text{TP} + \text{FP} +\text{FN})$ where $\text{FN}$ denotes the number of false negatives. The F1 score takes on values between $0$ and $1$, and 
a higher value indicates better support recovery.
Regarding the selection of the tuning parameter $\lambda$, we use five-fold cross-validation for all methods. The matrix parameter $\M B$ for the group GMC method is set according to \eqref{B-eqn}. Namely, given a convexity-preserving parameter $\alpha$, $\M B$ varies with $\lambda$, but we always have $\lambda \M B \T \M B = \alpha \M X \T \M X$ so that the convexity degree of the optimization problem is fixed.
We include the study of the ANOVA model in this section and relegate the investigation of the additive model to the supplement. We also report run times of different methods for all the simulation experiments in the supplement.

The data generation process of the ANOVA model is as follows. Four categorical variables $Z_1, Z_2, Z_3$ and $Z_4$ are generated from  a centered multivariate normal distribution with covariance between $Z_i$ and $Z_j$ being $\rho^{|i-j|}$ for $i,j=1, ...,4$. Then each $Z_i$ is trichotomized to $0, 1$ or $2$ if it is smaller than $\Phi^{-1}(\frac{1}{3})$, larger than $\Phi^{-1}(\frac{2}{3})$ or in between, where $\Phi$ is the cumulative density function (CDF) of the standard normal  distribution.  The true regression model is
\begin{multline}
\label{M2}
      y \amp = \amp 3\mathbbm{1}(Z_1=1)+2\mathbbm{1}(Z_1=0)+3\mathbbm{1}(Z_2=1)+
    2\mathbbm{1}(Z_2=0)+\mathbbm{1}(Z_1=1,Z_2=1)
    +\\~~~~~1.5 \mathbbm{1}(Z_1=1,Z_2=0)+2\mathbbm{1}(Z_1=0,Z_2=1)+2.5 \mathbbm{1}(Z_1=0,Z_2=0)+\epsilon,
\end{multline}
where $\epsilon$  is normally  distributed with mean zero and variance $\sigma^2$. Therefore, we have $32$ covariate variables from ten groups, where four of them with a group size of two represent the main
effects and the other six groups with a group size of four indicate the two-way interactions. The response variable, however, is only related to three groups of covariates as shown in model \eqref{M2}.  We next consider three different cases to explore the possible effects of interesting factors.

\begin{enumerate}
    \item [\textbf{\textit{Case C1:}}] The first factor we are interested in is the SNR, defined as $\lVert \M X \V \beta^\star \lVert_2/(\sqrt{n}\sigma)$. We consider uncorrelated groups, namely $\rho=0$ in the data generation process. 
    We set a sequence of $\sigma$ so that the SNR ranges from $1$ to $5$. The sample size $n$ is fixed as $100$ for each setting. To better understand  how the convexity-preserving parameter $\alpha$ affects the performance of the proposed group GMC method, we report the results of the group GMC with  $\alpha \in \{0.2, 0.4, 0.6, 0.8, 1\}$. These results also provide guidance on how to set $\alpha$  for the group GMC in practice.  
\end{enumerate}

\begin{figure}
    \centering
    \includegraphics[width=1\columnwidth]{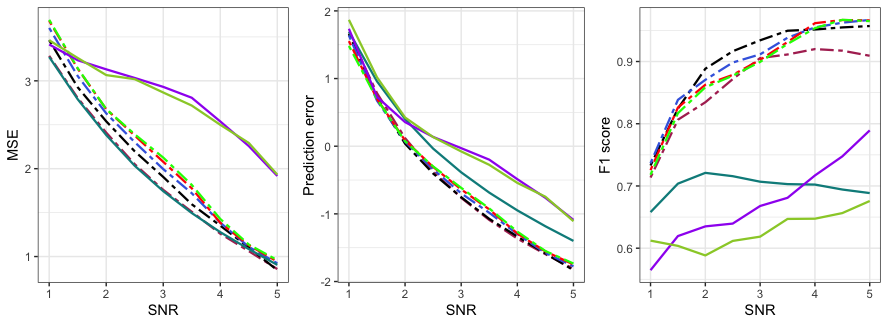}\\
     \includegraphics[width=0.8\columnwidth]{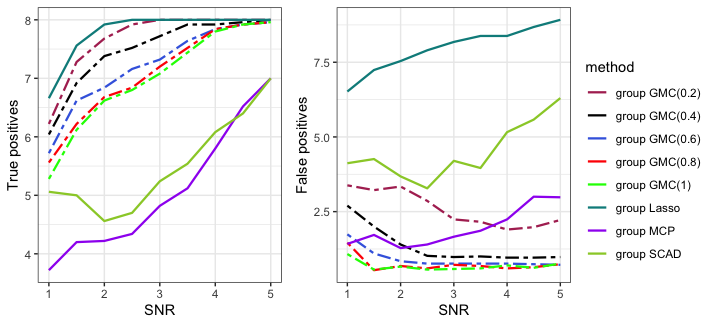}
    \caption{Results for \textit{Case C1}: Impact of SNR. Group  GMC($\cdot$) stands for the group GMC with a specific value of $\alpha$. Average performance based on $100$ simulation replicates for each method. MSE and Prediction error are on a log scale.}
    \label{fig-M2-SNR}
\end{figure}

  Figure \ref{fig-M2-SNR} presents the impact  of SNR on the performance of the four methods for model \eqref{M2}. As expected, the MSE of the estimated coefficients and the prediction error decrease as the SNR increases for all methods. 
  The group Lasso achieves the  lowest MSE, while the group GMC  gives the lowest prediction error among the four methods. The convexity-preserving parameter  $\alpha$ does not show a significant effect on the prediction performance of the group GMC, but it has a mild effect on the coefficient estimation. As indicated in the top left panel of Figure \ref{fig-M2-SNR}, a smaller value of $\alpha$ leads to a lower MSE.
  When it comes to support recovery, the group GMC shows a distinct advantage  over the other three methods. It achieves a higher F1 score than existing methods in all SNR settings. 
 The two plots on the bottom
panel  of Figure \ref{fig-M2-SNR} display the variable selection results of different methods in detail. The group Lasso obtains the most true positives but also the most false positives. Both group SCAD and group MCP miss some true positives and also include some irrelevant variables into the ANOVA model. The group GMC, however, can achieve a number of true positives comparable with the group Lasso while maintaining its number of false positives at a very low level.
  The convexity-preserving parameter  $\alpha$ indeed affects the variable selection performance of the group GMC. Both numbers of  true and false positives decrease as the value of $\alpha$ increases. In other words, a large value of  $\alpha$ in the  group GMC results in a sparse model.
 In general, a range of $0.4 < \alpha <1$ works well for this ANOVA example.

\begin{enumerate}
    \item [\textbf{\textit{Case C2:}}] When comparing different grouped variable selection methods, one factor of interest is to what extent the correlation among groups impacts their performance. For that purpose, we set a grid of values for $\rho$, $\rho=\{0, 0.2, 0.4, 0.6, 0.8\}$, so that the correlation between $Z_i$ and $Z_j$ is $\rho^{|i-j|}$ for $i \neq j$. We fix the SNR of the regression model to be $2$ and the sample size to be $100$ for each run. We set the convexity-preserving parameter $\alpha=0.6$ for the group GMC method.
\end{enumerate}

\begin{figure}
    \centering
    \includegraphics[width=1\columnwidth]{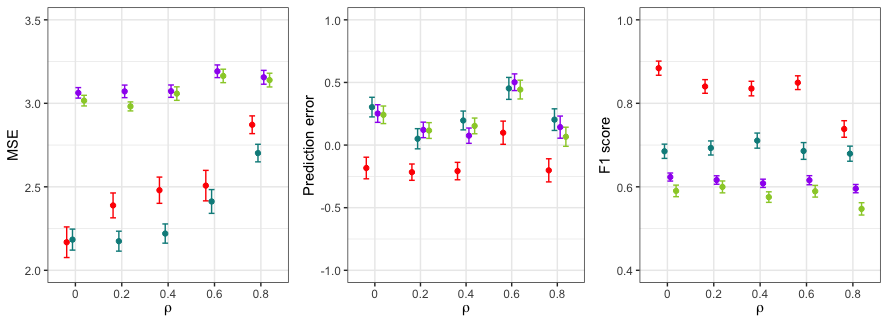}\\
     \includegraphics[width=0.8\columnwidth]{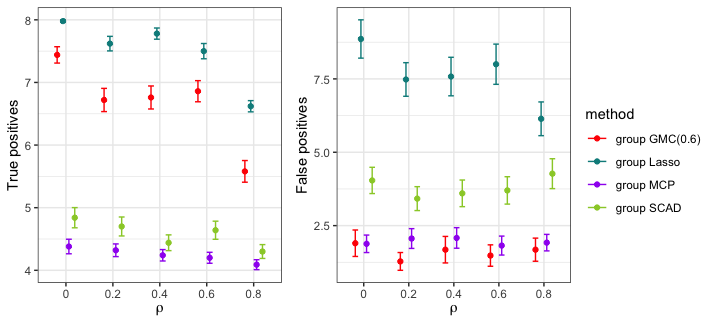}
    \caption{Results for \textit{Case C2}: Impact of group correlation. Average performance plus/minus one standard error based on $100$ simulation replicates for each method. MSE and Prediction error are on a log scale.}
    \label{fig-M2-cor}
\end{figure}

 Figure \ref{fig-M2-cor} shows the performance of the four methods under different group correlations. Both the group GMC and group Lasso produce worse estimation as the correlation $\rho$ increases, while the group MCP and group SCAD fail to achieve comparable estimation even in the uncorrelated setting. For the model prediction, all four methods are  relatively stable across different correlation settings, while the group GMC compares favorably with the other three. Regarding the variable selection with respect to the F1 score, group GMC visibly outperforms the existing three methods. All methods see a drop in F1 score when the correlation $\rho$ reaches up to $0.8$. The plots of true and false positives provide detailed insight into the variable selection performance of different methods. 
 The group Lasso includes the most false positives into the model, although it leads others in the inclusion of true positives. In contrast, the group MCP and group SCAD build much sparser models, thus missing some true positives. The group GMC is capable of obtaining true positives comparable with the group Lasso and excluding those irrelevant variables from the regression model. When $\rho =0.8$, all methods suffer a drop in their true positives, resulting in the drop in their F1 scores as seen in the corresponding plot.

\begin{enumerate} 

\item [\textbf{\textit{Case C3:}}]  In this third case, our goal is to explore the impact of the problem dimension on the performance of different methods. To that end, we set three different dimension settings where four, ten, and sixteen independent categorical variables $Z_i$ are generated accordingly in each setting and then trichotomized in the same way as described above. As a result, the problem dimension $p$ is $32$, $200$, and  $512$, respectively. But the response variable $y$ remains generated according to  model \eqref{M2} with an SNR of $2$, namely the number of true positives is $8$ in all dimension settings. The sample size is $100$ for each setting. We again fix the convexity-preserving  parameter $\alpha$ of the group GMC as $0.6$.
\end{enumerate}

\begin{figure}
    \centering
    \includegraphics[width=1\columnwidth]{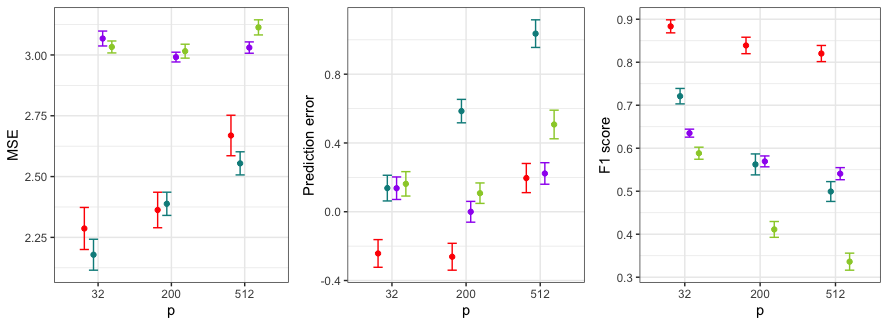}\\
     \includegraphics[width=0.8\columnwidth]{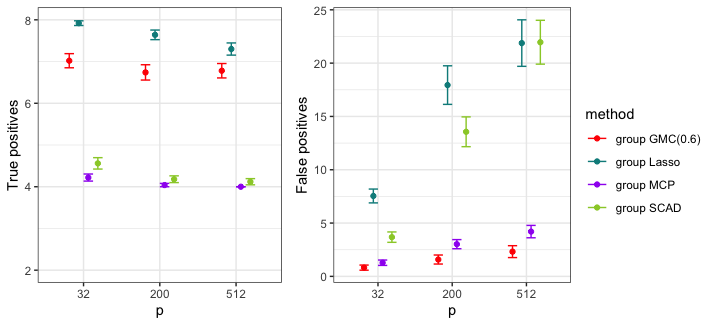}
    \caption{Results for \textit{Case C3}: Impact of problem dimension. Average performance plus/minus one standard error based on $100$ simulation replicates for each method. MSE and Prediction error are on a log scale.}
    \label{fig-M2-dim}
\end{figure}

Figure  \ref{fig-M2-dim} summarizes the simulation results. In terms of the coefficient estimation, the group MCP  and group SCAD behave quite similarly and much worse than the group Lasso and group GMC. Regarding the model prediction, 
the group GMC fares well in all dimension settings. 
With respect to variable selection, the group GMC shows a distinct advantage  over the existing three methods across different problem dimensions thanks to its robust behavior
 with respect to both true and false positives. 
On the one hand, while group MCP performs well at excluding false positives, it errs on the side of being too conservative and misses some true positives. On the other hand, the group Lasso and group SCAD, however, select too many irrelevant variables into the regression model, especially for the high-dimensional scenarios. 

\section{Real Data Application}
\label{sec6}

We apply our group GMC method on the birth weight data set from \cite{birthwt},
which studies risk factors associated with low infant birth weight.
The data set is publicly available in the R package \texttt{MASS} and contains $189$ observations of one response variable (infant birth weight) and  eight explanatory variables from the mother, including both continuous and categorical factors. 
We include detailed  descriptions of the data set in  Table \ref{data}.
As with \cite{grouplasso}, we take into account the preliminary analysis that both mother’s age and weight have  non-linear effects on the birth weight. Therefore, we model these two effects by third-order polynomials. Finally, we get sixteen predictors from eight groups to fit a linear regression model.

Following our simulation studies,  we analyze the data using  the proposed group GMC as  well as the  group Lasso,  group MCP, and group SCAD. For group GMC, we again set the matrix parameter $\M B$ according to \eqref{B-eqn} and choose  $\alpha=0.8$ based on our experience from the simulation studies. For evaluation, we first  randomly sample three-quarters of the observations ($142$ cases) as a training set for selecting the tuning parameter $\lambda$ by ten-fold cross-validation. Then we use the obtained  tuning parameter to fit the full data to get  the estimated coefficients. Finally, we compute  the prediction error based on the  testing set of the remaining one-quarter records. 

\begin{table}
\def~{\hphantom{0}}
\caption{Description of the birth weight data set}
{%
\begin{tabular}{p{4cm}p{3cm}p{10cm}}
\\
Name & Type & Variable  description \\[1.5ex]
 Birth weight & Continuous & Infant  birth weight in kilograms\\[0.7ex]
  Mother's age & Continuous & Mother's age in years\\[0.7ex]
  Mother's weight & Continuous & Mother's weight in pounds at last menstrual period\\[0.7ex]
 Race & Categorical & Mother's race (white, black or other)\\[0.7ex]
 Smoking & Categorical & Smoking status during pregnancy (yes or no)\\[0.7ex]
 \# Premature & Categorical & Previous premature labors (0, 1, or more)\\[0.7ex]
 Hypertension & Categorical & History of hypertension (yes or no)\\[0.7ex]
 Uterine irritability & Categorical & Presence of uterine irritability (yes or no)\\[0.7ex]
 \# Phys. visits & Categorical & Number of physician visits during the first trimester ($0, 1, 2,$ or more)\\[0.7ex]
\end{tabular}}
\label{data}
\end{table}

\begin{table}
\def~{\hphantom{0}}
\caption{Summarized results for the birth weight data}
{%
\begin{tabular}{lccc}
\\
 & Prediction error & \# nonzero groups & Excluded groups\\[1.5ex]
 Group Lasso & 0.36 & 8 & none\\[0.7ex]
 Group SCAD & 0.35 & 8 & none\\[0.7ex]
 Group MCP & 0.35 & 7 & \# Phys. visits\\[0.7ex]
 Group GMC & 0.35 & 7 & \# Phys. visits\\[0.7ex]
\end{tabular}}
\label{Tab1}
\end{table}

Table \ref{Tab1} summarizes the prediction errors, number of nonzero groups, and the excluded groups from the four methods. The group Lasso and group SCAD fail to exclude any group from the model. Both group MCP and group GMC, however, regard the number of  physician
visits during the first trimester as an unimportant factor to the infant birth weight. The prediction errors obtained from the four different methods are comparable.

\begin{figure}
      \includegraphics[ width=0.48\textwidth]{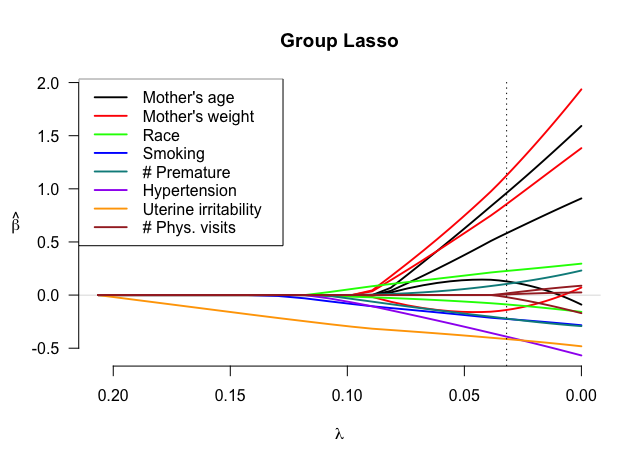}
\hspace{\fill}
     \includegraphics[ width=0.48\textwidth]{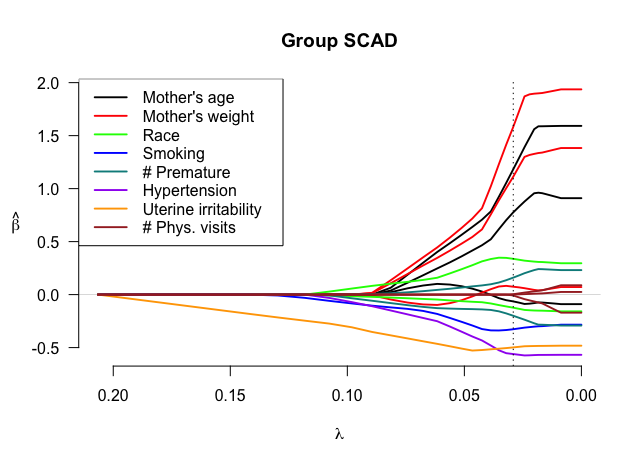}
     
      \includegraphics[ width=0.48\textwidth]{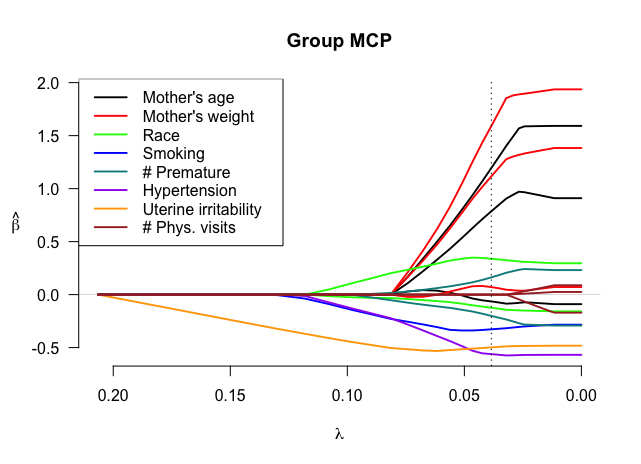}
\hspace{\fill}
     \includegraphics[ width=0.48\textwidth]{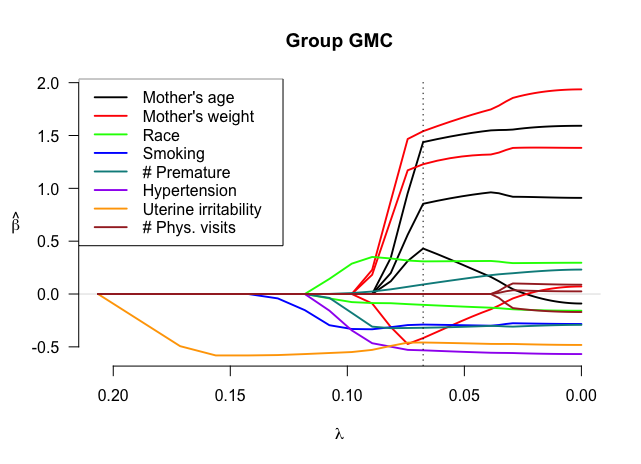}
\caption{Solution paths of the birth weight data from four different methods. The dotted vertical line in each subplot indicates the selected $\lambda$ via ten-fold cross-validation.}
\label{fig-path}
\end{figure}{}

It seems that, for this birth weight data analysis, the group GMC does not exhibit any advantage over existing methods. Nevertheless, the solution paths from the four methods, as shown in Figure \ref{fig-path}, tell a different story.  The estimated coefficients from the group GMC, as indicated by the vertical dotted line, undergo noticeably less shrinkage than those of the group Lasso and are similar to the estimates from the group MCP and group SCAD. This confirms the unbiased (or at least less biased) estimation of the group GMC as a nonconvex penalization method. What is more, the group GMC method is more robust against the tuning parameter selection compared to the other three methods. It is anticipated that the estimated coefficients are increasingly shrunk as $\lambda$ increases, and thus fewer variables are selected into the regression model.  But as shown in Figure \ref{fig-path}, the estimated coefficients and selected variables are stable over $\lambda \in [0.04, 0.07]$ for the group GMC, while the other three methods do not have as comparably a wide range of $\lambda$.
This insensitivity furnishes some evidence that the group GMC method can potentially blunt estimation bias successfully while still simultaneously achieving satisfactory variable selection.

\section{Discussion}
\label{sec7}

In this paper, we used the convex-nonconvex strategy to  propose a novel concave penalty, called the group  GMC, for grouped variable selection and coefficient estimation in linear regression. The group GMC penalty is a variant of  the GMC penalty and thus inherits its characteristic that it is  able to maintain the convexity of the corresponding optimization problem. Therefore, the group GMC eliminates the possibility of suboptimal local minima while maintaining unbiased estimation as a nonconvex penalization approach.
We formulated linear regression with the group GMC penalization as a convex optimization problem, or more specifically a saddle-point problem when a certain condition is satisfied. The resulting group GMC estimator enjoys desirable properties which help accelerate  numerical computation and tuning parameter selection. Our algorithm for computing the group GMC estimator is guaranteed to converge to the global minimizer of the group GMC optimization problem. Additionally, we analyzed statistical properties of the group GMC estimator as well as the original GMC estimator. Our results are the first to establish the $l_2$-norm error bounds for the GMC least squares estimators and as such, provide novel insights about the performance of convex nonconvex penalization. 
In our simulation study, we  compared the practical performance of the group GMC with the group Lasso, group MCP, and group SCAD via a comprehensive evaluation, including variable selection, coefficient estimation, and model prediction. 
Through a battery of simulation experiments, we found that the group GMC can achieve better or at least competitive performance in comparison with the existing three methods under different scenarios such as different SNRs, correlated or uncorrelated groups, and different dimension settings. A real data application  displays the advantage of the group GMC method in its robustness in unbiased coefficient estimation  and grouped variable selection.

Several related studies can be done in the future. First of all, how to set the matrix parameter $\M B$ warrants more exploration and investigation. We discussed how $\M B$ could affect the error bound of the group GMC estimator in Section \ref{sec4} but anticipate that other approaches to set $\M B$ could further improve the performance of the group GMC, both theoretically and practically.  
Second, the group GMC method could  be extended to generalized linear  models to deal with  grouped variable selection  problems in other high-dimensional cases. More generally, the convex nonconvex strategy could be applied to other sparse learning scenarios so that one can  enjoy the advantages of convex optimization and nonconvex penalization simultaneously.



\bigskip
\begin{center}
{\large\bf SUPPLEMENTARY MATERIAL}
\end{center}
The supplementary material consists of technical proofs, details of the adaptive PDHG algorithm, and additional simulation experiments on an additive model.

\bibliographystyle{asa}

\bibliography{paper_ref}
\end{document}